\documentclass[%
 reprint,
nofootinbib,
 amsmath,amssymb,
 aps,
prd,
showkeys,
]{revtex4-2}

\usepackage{graphicx}
\usepackage{dcolumn}
\usepackage{bm}
\usepackage{hyperref}
\usepackage{orcidlink}
\hypersetup{
    colorlinks=true,
    linkcolor=blue,
    filecolor=magenta,      
    urlcolor=cyan,
    pdfpagemode=FullScreen,
}
\usepackage{color,soul}
\usepackage{comment}
\usepackage[normalem]{ulem}

\usepackage{soul}
\usepackage{color}
\usepackage{slashed}

\newcommand{\be}{\begin{equation}}
\newcommand{\ee}{\end{equation}}

\begin{document}

\title{Scattering and Femtoscopic Correlation Functions of the $\Sigma_c^{++}\pi^{+}$, $\Sigma_c^{0}\pi^{-}$ and $\Sigma_b^{+}\pi^{+}$ Systems}

\newcommand{\ific}{Instituto de F\'{\i}sica Corpuscular (centro mixto CSIC-UV),
Institutos de Investigaci\'on de Paterna,
C/Catedr\'atico Jos\'e Beltr\'an 2, E-46980 Paterna, Valencia, Spain}
\newcommand{\ice}{Institute of Space Sciences (ICE, CSIC), Campus UAB,  Carrer de Can Magrans, 08193 Barcelona, Spain}
\newcommand{\ieec}{Institut d'Estudis Espacials de Catalunya (IEEC), 08860 Castelldefels (Barcelona), Spain}
\newcommand{\teoUV}{
Departamento de F\'\i sica Te\'orica and IFIC, Centro Mixto Universidad de Valencia-CSIC,
Institutos de Investigaci\'on de Paterna, Aptdo. 22085, E-46071 Valencia, Spain}

\author{Mikel F. Barbat\orcidlink{0009-0007-4357-5823}}
 \email{mikel.fernandez@ific.uv.es}
  \affiliation{\teoUV}
 \author{Juan Nieves\orcidlink{0000-0002-2518-4606}}
\email{jmnieves@ific.uv.es}
\affiliation{\ific}
\author{Laura Tolos\orcidlink{0000-0003-2304-7496}}
\email{tolos@ice.csic.es}
 \affiliation{\ice}
  \affiliation{\ieec}

\date{\today}

\begin{abstract}
We present predictions for scattering observables and femtoscopic correlation functions (CFs) of the $I=2$ $\Sigma_c^{++}\pi^{+}$, $\Sigma_c^{0}\pi^{-}$ systems and its heavy-flavor counterpart $\Sigma_b^{+}\pi^{+}$. In both heavy-quark sectors, the strong interaction is formulated within two distinct theoretical frameworks, each constrained to reproduce the lowest-lying odd-parity isoscalar spin-$1/2$ resonances, $\Lambda_c(2595)$ and $\Lambda_b(5912)$, respectively. While the $\Sigma_c^{0}\pi^{-}$ pair is governed solely by the strong interaction, electrostatic contributions are included in the other two channels involving charged particles through relativistic Coulomb wave functions. We show that the differences observed in the scattering observables between the two strong-interaction models arise mainly from the specific ultraviolet regularization schemes employed. The inclusion of Coulomb effects induces only a very small increase in both the scattering length and the effective range. The resulting CFs in the charm and bottom sectors display analogous global features, in agreement with expectations from heavy-quark flavor symmetry. Both, the $\Sigma_c^{++}\pi^+$  and $\Sigma_b^{+}\pi^{+}$ CFs, when computed including only the strong interaction, exhibits substantial discriminating power among the different models. However, once Coulomb effects are incorporated, the CFs become largely affected by the repulsive electrostatic interaction, which diminishes their sensitivity to the details of the underlying strong dynamics, thereby reducing the capability to differentiate between theoretical descriptions. Thus, the $\Sigma_c^{0}\pi^{-}$ CF—being free from Coulomb effects—provides the most suitable observable for constraining the strong dynamics of the isotensor $\Sigma_c\pi$ system. 
\end{abstract}

\maketitle

\section{Introduction}

Quantum Chromodynamics (QCD) constitutes the fundamental theory of the strong interaction. Perturbative techniques can successfully describe processes at high energies, corresponding to distance scales shorter than the nucleon size. In contrast, in the low-energy regime QCD becomes strongly coupled and hadrons emerge as the relevant degrees of freedom. Their mutual interactions are not yet fully constrained and remain challenging to access experimentally.

The unstable nature of hadrons containing charm or bottom quarks prevents the use of traditional scattering experiments as a feasible method to study their interaction. Under these conditions, femtoscopy provides a powerful tool to investigate interactions among hadrons (see the review of Ref.~\cite{Fabbietti:2020bfg}). This technique consists in determining the correlations in momentum space for the hadron-hadron pair produced in high-multiplicity collisions. Experimentally, the correlation function (CF) is constructed as the ratio of the relative momentum distribution of pairs produced in the same event with a reference distribution obtained from mixed events~\cite{HADES:2016dyd,Shapoval:2014yha}. Theoretically, the CF can be computed from the product of the source function, describing the probability distribution that the two particles from the pair are emitted at a relative distance, multiplied by the squared two-body wave function of the hadron pair~\cite{Koonin:1977fh,Pratt:1990zq}.

The ALICE Collaboration at LHC has carried out an extensive program on the study of hadron-hadron interactions using femtoscopic techniques~\cite{ALICE:2017jto,ALICE:2017iga,ALICE:2018nnl,ALICE:2018ysd,ALICE:2019igo,ALICE:2019eol,ALICE:2019gcn,ALICE:2019hdt,ALICE:2019buq,ALICE:2020wvi,ALICE:2020mfd,ALICE:2020mkb,ALICE:2021njx,ALICE:2021ovd,ALICE:2021cyj,ALICE:2021cpv,ALICE:2022yyh,ALICE:2022uso,ALICE:2022mxo,ALICE:2023wjz,ALICE:2023gxp,ALICE:2023zbh,ALICE:2023eyl,ALICE:2023sjd,ALICE:2025aur,ALICE:2025wuy}, and recently this method has been extended to systems involving charm quarks. Up to now, these latter analyses have focused on the study of the $D$ meson through the CFs of $D^{(\pm)} \pi^{(\pm)}$ and $D^{(\pm)} K^{(\pm)}$ \cite{ALICE:2024bhk}, as well as $D^- p$ and $D^+ \bar p$ \cite{ALICE:2022enj}, measured in high-multiplicity $pp$ collisions at $\sqrt{s}=13$ TeV.

In the charm sector, the correlation function is obtained through the reconstruction of charmed hadrons via their hadronic decay channels. This requires excellent particle identification, as well as precise tracking and spatial resolution for charged particles and decay vertices. Experiments at the LHC, such as ALICE at CERN, provide these capabilities through high-resolution silicon detectors and large-acceptance tracking systems, enabling the reconstruction of charmed hadrons with sufficient precision. Combined with the large data samples collected in high multiplicity $pp$ and heavy-ion collisions, these conditions make femtoscopic studies involving charm feasible \cite{ALICE:2025cdf,ALICE:2025ygv}.

Parallel to the experimental developments, several theoretical studies have analyzed charmed hadrons using femtoscopic techniques. These include investigations of hidden charmed states, such as the $X(3872)$~\cite{Kamiya:2022thy,Abreu:2025jqy}, $T_{cc}$~\cite{Kamiya:2022thy,Vidana:2023olz,Albaladejo:2024lam}, $Z_c(3900)$ and $Z_{cs}(3985)$~\cite{Liu:2024nac}, as well as $D$ mesons with baryons~\cite{Liu:2023wfo,Barbat:2025orm}. Other works have analyzed the CF of open-charm mesons with light mesons \cite{Albaladejo:2023pzq,Torres-Rincon:2023qll,Liu:2023uly,Ikeno:2023ojl,Khemchandani:2023xup}, as well as charmonium-nucleon \cite{Liu:2025oar} and $\alpha$-charmonium systems \cite{Etminan:2025tiy}.

Motivated by these previous studies, we analyze in this work the CF of the $\Sigma_c^{++}\pi^+$ and $\Sigma_c^0\pi^-$ systems in the isospin $I=2$ tensor channel. As a first step, we neglect the Coulomb interaction and solve the Bethe–Salpeter equation (BSE) using as kernel $V$ (i.e., the two-particle irreducible amplitude), the $S$-wave leading-order Weinberg–Tomozawa (WT) chiral interaction. The resulting amplitudes fulfill exact elastic unitarity. We adopt on-shell–type approximations~\cite{Nieves:1999bx}, in which the contributions of the off-shell components of the interaction are effectively accounted by some low-energy constants or modeled by separable ans\"atze for the potential that parametrize the ultraviolet (UV) behavior, namely $V(q,q')\to V(k,k)\,F_{\rm UV}(q)\,F_{\rm UV}(q')$, with $k$ the on-shell momentum and  $F_{\rm UV}(q)$ a form factor that renders the BSE amplitude finite. As a consequence, the unitarized amplitudes depend exclusively on a single UV cutoff, $\Lambda$, introduced through $F_{\rm UV}$ to renormalize the BSE. 

Within this framework, we fix the UV cutoff in the $I=0$ $\Sigma_c \pi$ sector, where it is constrained by the properties of the odd-parity $\Lambda_c(2595)$ resonance. In the isoscalar channel, two strong-interaction potential models are commonly considered.

The first approach—developed in Refs.~\cite{Hofmann:2005sw,Mizutani:2006vq,Jimenez-Tejero:2009cyn}—extends the WT interaction in the $\Sigma_c \pi$ channel by incorporating additional channels, most notably $ND$, and implements the corresponding coupled-channel dynamics within an SU(4)-symmetry–based framework. The second~\cite{Nieves:2024dcz} augments the chiral WT $\Sigma_c \pi$ interaction with an additional potential generated by the exchange of a constituent quark model (CQM) state, whose mass is found to lie very close~\cite{yoshida:2015tia} to the nominal mass of the $\Lambda_c(2595)$\footnote{ We note that in the $I=2$ $\Sigma_c\pi$ channel it is not possible to exchange a CQM-state and, therefore, the main difference between both models arises from the details of the renormalization procedure, as mentioned above.}.

Since the $\Sigma_c^{++}\pi^+$  system  involves charged hadrons, the inclusion of the Coulomb interaction in the calculation of the CF is also essential. We follow the formalism of Ref.~\cite{Albaladejo:2025kuv}, originally applied to the $pp$ CF, and extend it to the present case by employing relativistic Coulomb wave functions. 

Our goal is not only to provide a robust determination of the phase shifts, but also to assess whether future measurements of the $\Sigma_c^0\pi^-$  and $\Sigma_c^{++}\pi^+$ CFs could indirectly discriminate between the two strong-interaction scenarios. In particular, the models differ in their treatment of the isoscalar sector, incorporating either $ND$ dynamics or explicit constituent quark model (CQM) degrees of freedom. These distinct microscopic mechanisms entail different ultraviolet (UV) implementations, which in turn propagate to the isotensor ($I=2$) channel. Next, we investigate the heavy-flavor partner in the bottom sector, namely the $\Sigma_b^+ \pi^+$ system. In this case, the UV cutoff is constrained by the $\Lambda_b(5912)$ resonance in the $I=0$ channel, and we perform an analysis analogous to that carried out in the charm sector, emphasizing the similarities arising from the approximate heavy-quark flavor symmetry (HQFS) of QCD.

For the sake of brevity, throughout the rest of this work—up to the Summary section—in the charm sector we will refer only to the $\Sigma_c^{++}\pi^{+}$ pair, with the understanding that the predictions for the $\Sigma_c^{0}\pi^{-}$ channel correspond to those obtained for the $\Sigma_c^{++} \pi^+$ system when Coulomb effects are switched off.

The paper is organized as follows. In Sec.~\ref{sec:phase-length-range}, we provide the details for the theoretical models used to construct the interaction between $\Sigma_c^{++}$ and $\pi^+$, as well as the formalism to obtain the scattering amplitude, while describing the implementation of the Coulomb interaction. We also calculate the $\Sigma_c^{++} \pi^{+}$ strong phase shift, scattering length and effective range. In Sec.~\ref{sec:CorrelationFunction} we present the calculation of the $\Sigma_c^{++} \pi^+$ CF, showing our results including the Coulomb interaction. In Sec.~\ref{sec:CF_SbP}, we present the predictions for the strong phase shift and the CFs in the heavy-flavor bottom $\Sigma_b^+\pi^+$ sector, and compare them with the corresponding results in the charm sector. Finally, in Sec.~\ref{sec:Summary} we give a summary of this work.

\section{$\Sigma^{++}_c \pi^+$ phase shift, scattering length and effective range}
\label{sec:phase-length-range}

We start by presenting the theoretical models for the description of the $\Sigma_c^{++} \pi^+$ effective strong amplitude and then compute the scattering amplitude including both the strong and Coulomb interactions. The $\Sigma_c^{++} \pi^+$ phase shift, scattering length and effective range are subsequently calculated considering only the strong interaction as well as including the Coulomb effects.

\subsection{Interaction in the $I=0$ sector and UV cutoff}
\label{sec:i0}

\subsubsection{ $(\Sigma_c \pi-ND)$ SU(4)-WT scheme} 

We begin by summarizing the model developed in Refs.~\cite{Hofmann:2005sw,Mizutani:2006vq,Jimenez-Tejero:2009cyn}. This effective SU(4) scheme includes the scattering of all possible baryon-meson channels in the different sectors characterized by the charm ($C$), strangeness ($S$), spin ($J$) and isospin ($I$) quantum numbers. The interaction can be described either by the WT potential \cite{Mizutani:2006vq} or the zero-range  approximation \cite{Jimenez-Tejero:2009cyn}. In the present work, we employ the WT potential given by:
\begin{align}
    V_{ij, \ \textrm{WT}}^{ISC} (s) &= - \dfrac{N}{4f_{\pi}^2}C_{ij}^{\textrm{WT}} (2\sqrt{s} - M_i -M_j),
    \label{eq:VWT}
\end{align}
where
\begin{equation}
    N= \sqrt{\dfrac{E_i^{\rm on}+M_i}{2M_i}} \sqrt{\dfrac{E_j^{\rm on}+M_j}{2M_j}},
\end{equation}
is a normalization factor arising from the Dirac spinors, $s$ stands for the baryon–meson Mandelstam variable, $M_i$ and $m_i$ are the baryon and meson masses of the channel $i$, whereas $E_i^{\rm on}=(s-m_i^2+M_i^2)/2\sqrt{s}$ is the center-of-mass (CM) energy of the baryon in the same channel. The coefficients $C_{ij}^{\textrm{WT}}$ are the matrix elements for the different $(C,S,J,I)$ sectors, whereas the $f_\pi= 92~\mathrm{MeV}$ is the pion decay constant. For $(C=1,S=0,J=1/2,I=0)$ the baryon-meson channels are $\pi \Sigma_c, DN, \eta \Lambda_c, K\Xi_c, K\Xi'_c, D_s \Lambda, \eta' \Lambda_c$, and the corresponding coefficients $C_{ij}^{\textrm{WT}}$ can be found in Ref.~\cite{Mizutani:2006vq}.

In order to obtain the effective strong amplitude, we solve the on-shell coupled-channel BSE, with the interaction kernel shown in Eq.~(\ref{eq:VWT}). The $S-$wave effective amplitude in the on-shell scheme can be written as

\begin{equation}
    T(s)=\dfrac{V(s)}{1-V(s)G(s)},
\end{equation}
with the matrix $G(s)$ constructed from the baryon-meson loop functions as, 
\begin{equation}
    G_i(s) = \int \dfrac{d^3q}{(2\pi)^3} \dfrac{E_i+\omega_i}{2 E_i \omega_i} \dfrac{2M_i }{s-(E_i+\omega_i)^2 + \mathrm{i} \epsilon} ,
    \label{eq:gloop} 
\end{equation}
Here, $E_i$ and $\omega_i$ denote the relativistic energies of the baryon and meson in channel $i$ with off-shell momentum $\vec{q}$. The loop function is evaluated using the analytical expression with a sharp-cutoff regularization, $F_{\rm UV}(q) = H(\Lambda - q)$  where $H$ is the Heaviside step function, as detailed in Ref.~\cite{Oller:1998hw}.

As extensively discussed in the literature, the $\Lambda_c(2595)$ state with spin-parity $J^P=1/2^-$, which mass is located precisely just below the $\Sigma_c\pi$ threshold, in the ($C=1,S=0,J=1/2,I=0$) sector can be generated dynamically through the baryon-meson scattering, with its physical mass obtained adjusting the UV regulator. In our case, we find a value for the cutoff of $\Lambda=653 ~ \mathrm{MeV}$. 

The same cutoff value will be employed in the analysis of the ($C=1, S=0, J=1/2, I=2$) sector, where only the $\Sigma_c^{++} \pi^+$ channel contributes. In this case, the WT coefficient is $C_{\Sigma_c\pi; I=2}^{\textrm{WT}} = 2$, which leads to a repulsive interaction. We note that, in contrast, the corresponding coefficient in the isoscalar sector is $C_{\Sigma_c\pi; I=0}^{\textrm{WT}} =  -4$, giving rise to an attractive interaction twice as strong in magnitude.

\subsubsection{ $\Sigma_c\pi$ [WT $\&$ CQM] scheme}

In Ref.~\cite{Nieves:2024dcz}, it is argued that the presence of a CQM state near 2620 MeV, as predicted in Ref.~\cite{yoshida:2015tia}, and its interplay with the isoscalar $\Sigma_c\pi$ channel should significantly influence the dynamics of the $\Lambda_c(2595)$. In principle, this effect is expected to be more pronounced than that of the $ND$ channel, whose threshold lies more than 200 MeV higher.
This hypothesis is further supported by the properties of its heavy-quark spin symmetry partner, the $\Lambda_c(2625)$ with $J^P=3/2^-$, as well as by analogous findings in the bottom sector, which includes the $\Lambda_b(5912)$ and $\Lambda_b(5920)$ resonances with $J^P=1/2^-$ and $J^P=3/2^-$, respectively. Further details can be found in the discussion of Fig.~1.1 of Ref.~\cite{Nieves:2024dcz}. 

In Ref.~\cite{Nieves:2024dcz}, $\Sigma_c\pi$ is treated as the only relevant channel also in the $\Lambda_c(2595)$ isoscalar sector, with the corresponding interaction being constructed by supplementing the chiral WT potential with an additional $\Sigma_c\pi\to \Sigma_c\pi$ contribution (see Fig.~2.1 of Ref.~\cite{Nieves:2024dcz}) arising from the exchange of an intermediate CQM state~\cite{Cincioglu:2016fkm,Nieves:2024dcz}, 
\begin{equation}
    V_{ex}(s)=2\mathring{M}_{CQM} \dfrac{d_c^2}{s-(\mathring{M}_{\rm CQM})^2},
    \label{eq:VCQM}
\end{equation}
where $d_c$ is a dimensionless parameter that controls the coupling strength of the CQM vertex and the bare mass of the CQM state, $\mathring{M}_{\rm CQM}=2628~\rm{MeV}$, taken from the result obtained for the  $I=0$ configuration in the CQM study carried out in Ref.~\cite{yoshida:2015tia}. Additionally for this model, we follow the renormalization procedure outlined in Refs.~\cite{Nieves:2024dcz,Nieves:2001wt}, where the $\Sigma_c\pi$ loop function is decomposed into a finite and a UV-divergent part
\begin{equation}
    G_{\Sigma_c\pi}(s)=G_{\Sigma_c\pi}(s_{\rm th})+\overline{G}_{\Sigma_c\pi}(s),
    \label{eq:RenormSub}
\end{equation}
where $s_{\rm th}=(M_{\Sigma_c}+m_\pi)^2$. The finite part of the loop function, $\overline{G}_i(s)$, can be written as
\begin{eqnarray}
\label{eq:RenormSub1}
        &&\overline{G}_{\Sigma_c\pi}(s)=\dfrac{2M_{\Sigma_c}}{(4\pi)^2}\times  \nonumber\\
        &&\left( \left[ \dfrac{M_{\Sigma_c}^2-m_\pi^2}{s}-\dfrac{M_{\Sigma_c}-m_\pi}{M_{\Sigma_c}+m_\pi} \right]\log{\dfrac{M_{\Sigma_c}}{m_\pi}}
         +L(s)\right) \label{eq:Gbar}
\end{eqnarray}
where the multi-valued function $L(s)$ is given by Eq.~(A10) of Ref.~\cite{Nieves:2001wt}. The imaginary part is
\begin{equation}
  {\rm Im}\left[\,\overline{G}_{\Sigma_c\pi}(s)\right]= -\frac{M_{\Sigma_c}}{4\pi\sqrt{s}}\,p  
\end{equation}
with the CM momentum $p=\lambda^{1/2}(s,M_{\Sigma_c}^2,m_\pi^2)/(2\sqrt{s})$, where $\lambda(x,y,z)=x^2+y^2+z^2-2xy-2yz-2xz$ is the K\"allen function. The divergent contribution of the loop function, $G(s_{\rm th})$, needs to be regularized, in our case using a sharp cutoff regulator, that can be computed analytically as
\begin{eqnarray}
\label{eq:RenormSub2}
    &&G^\Lambda_{\Sigma_c\pi}(s_{\rm th})=\dfrac{1}{4\pi^2} \dfrac{M_{\Sigma_c}}{m_\pi+M_{\Sigma_c}}\left( m_\pi \ln{\dfrac{m_\pi}{\Lambda+\sqrt{\Lambda^2+m_\pi^2}}} \right. \nonumber \\ && \left. +M_{\Sigma_c} \ln{\dfrac{M_{\Sigma_c}}{\Lambda+\sqrt{\Lambda^2+M_{\Sigma_c}^2}}} \right). 
\end{eqnarray}
Note that $G^\Lambda_{\Sigma_c\pi}(s_{\rm th})$ should be regarded as an effective low-energy constant that effectively accounts for unknown short-distance dynamics and higher-order contributions. Its value is fixed indirectly through the UV cutoff $\Lambda$, which should be adjusted to reproduce the on-shell amplitude near threshold.

The solution of the BSE with this kernel may generate a pole in the first Riemann sheet of the $\Sigma_c\pi$ amplitude. Following Ref.~\cite{Nieves:2024dcz}, the parameters $d_c$ and the UV cutoff are determined by requiring this pole to coincide with the physical mass of the $\Lambda_c(2595)$ resonance. This condition gives rise to a family of solutions characterized by pairs of $(d_c,\Lambda)$ values, each reproducing the physical $\Lambda_c(2595)$ mass. As a result, the model provides a set of possible effective strong amplitudes, depending on the selected coupling strength.

For the present work we take two limiting values from Ref.~\cite{Nieves:2024dcz} that reproduce the $\Lambda_c(2595)$ state, that is ($d_c=0.69$, $\Lambda=650~\mathrm{MeV}$) and ($d_c=1.15$, $\Lambda=400~\mathrm{MeV}$). We will use this range of values for the cutoff to compute the effective strong interaction in the $I=2$ $\Sigma \pi$ sector  within this model. 

\begin{figure}[h!]
    \centering
    \includegraphics[width=\linewidth]{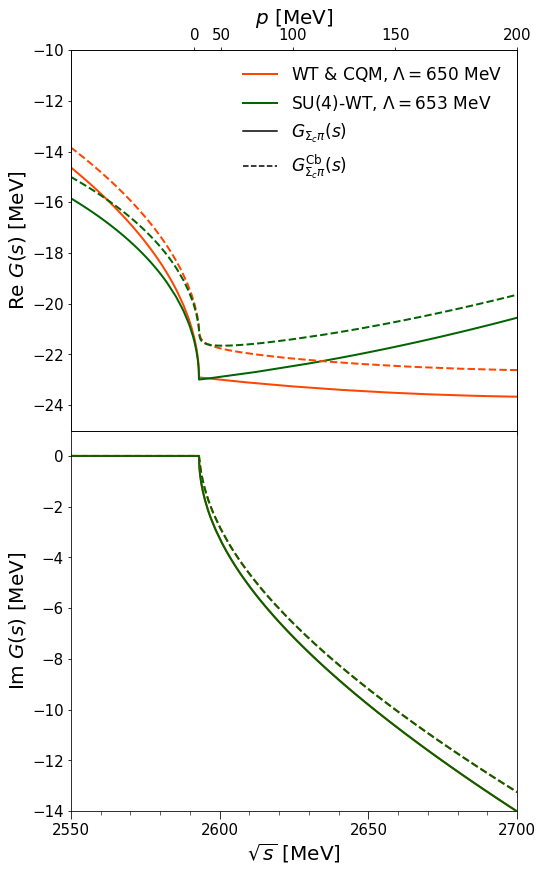}
    \caption{ Real (top) and imaginary (bottom) parts of $G_{\Sigma_c\pi}(s)$ as functions of the total CM energy $\sqrt{s}$ (lower x-axis) and the CM momentum $p$ (upper x-axis). The solid lines represent the loop functions obtained within the SU(4)-WT (green) and the $\Sigma_c\pi$ [WT $\&$ CQM] (orange) renormalization schemes without Coulomb interaction (denoted by $G_{\Sigma_c\pi}(s)$), while the dashed lines show the corresponding results including the repulsive Coulomb interaction (denoted by $G^{\mathrm{Cb}}_{\Sigma_c\pi}(s)$). We take the UV cutoff $\Lambda=650$ MeV in the $\Sigma_c\pi$ [WT $\&$ CQM] case, similar to the one obtained for the SU(4)-WT model ($\Lambda=653$ MeV). }
    \label{fig:Gs}
\end{figure}

The distinct renormalization schemes  adopted in the SU(4)-WT and the $\Sigma_c\pi$ [WT $\&$ CQM] approaches lead to different $\Sigma_c\pi$ loop functions, as shown in Fig.~\ref{fig:Gs} by the solid lines. In the former case, the full loop function $G_{\Sigma_c\pi}(s)$ is evaluated numerically by regulating the momentum integral with a cutoff $\Lambda$. In contrast, within the $\Sigma_c\pi$ [WT $\&$ CQM] scheme, only the logarithmically divergent part—namely, the loop function at threshold—is computed using the UV cutoff. This procedure avoids  introducing cutoff artifacts into the finite part, $\overline{G}_{\Sigma_c\pi}(s)$, which become apparent in the real part and grow as one moves away from threshold. There are no effects on the imaginary part because the CM moment remains below $\Lambda$.  For future reference, we show in the same figure with dashed lines the loop functions obtained with these two schemes, including Coulomb effects.

\subsection{Strong isotensor phase shifts in absence of Coulomb}
\label{sec:i2}

\begin{figure}[h!]
    \centering
    \includegraphics[width=\linewidth]{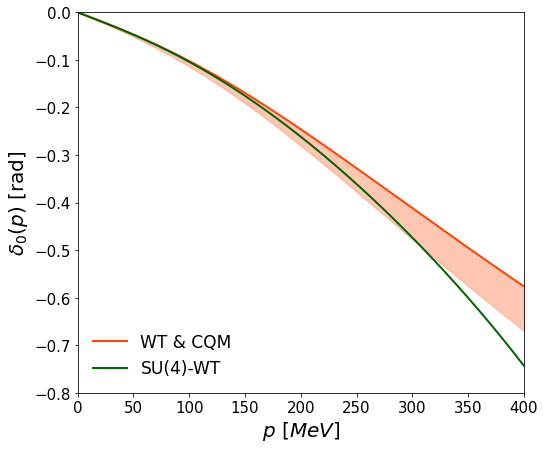}
    \caption{Strong isotensor $\Sigma_c\pi$ $S$-wave phase shift as a function of CM momentum. Orange (green) solid line: predictions of the $\Sigma_c\pi$ [WT $\&$ CQM] model with $\Lambda = 650~\mathrm{MeV}$ (SU(4)-WT model with $\Lambda = 653~\mathrm{MeV}$). Shaded band: phase shift range for the $\Sigma_c\pi$ [WT $\&$ CQM] model when the UV cutoff is varied from 400 to 650 MeV.}
    \label{fig:PS}
\end{figure}

Now we can compute the $\Sigma_c^{++} \pi^+$ strong ($I=2$) $S-$wave phase shift in the absence of the Coulomb interaction to compare the outcome from both theoretical approaches. The strong isotensor $\Sigma_c\pi \to \Sigma_c\pi$ phase shift $\delta(p)$ can be obtained from the solution $T(s)$ of the BSE\footnote{From now on, we omit the indices $i,j$ since for $I=2$ there is only one channel, $\Sigma_c^{++} \pi^+$.} as~\cite{Tolos:2007vh}
\begin{equation}
    f(p)=\dfrac{e^{2i\delta_0(p)}-1}{2ip} = -\dfrac{M_{\Sigma_c}}{4\pi\sqrt{s}}T(s),
\end{equation}
with $f(p)$ the standard scattering amplitude. These strong phase shift are shown here in Fig.~\ref{fig:PS}. For the $\Sigma_c\pi$[WT $\&$ CQM] approach, the two limiting cutoff values lead to a band of values for the phase shift, with the orange solid line indicating the case of $\Lambda=650$ MeV. We note that the two models exhibit similar behavior for low momenta. In the $I=2$ $\Sigma_c \pi$ sector, the SU(4)-WT and $\Sigma_c\pi$[WT $\&$ CQM] models share the same potential (as previously discussed, there is no possible CQM-state exchange for $I=2$). The differences therefore arise from the regularization scheme used for the loop function. As shown, this has a visible impact at momenta larger than 200 MeV. 
\\

\subsection{Inclusion of the Coulomb interaction}
We now proceed to include the Coulomb interaction in our computation of the scattering amplitude of the $\Sigma_c^{++} \pi^+$ system.  Within the non-relativistic Schr\"odinger (or Lippmann-Schwinger)  framework, when the Coulomb interaction is included, the full amplitude can be decomposed into the pure Coulomb contribution and a strong amplitude modified by the electromagnetic interaction (Gell-Mann–Goldberger decomposition \cite{Gell-Mann:1953dcn,Zorbas:1976cd}):
\begin{equation}
    T(\vec{p},\vec{p}^{\, \prime})=T_C(\vec{p},\vec{p}^{\, \prime})+T_{SC}(\vec{p},\vec{p}^{\, \prime}).
\end{equation}
with $\vec{p}$ ($\vec{p}^{,\prime}$) denoting the incoming (outgoing) three-momentum of the $\Sigma_c\pi$ system in the CM frame, $T_C(\vec{p},\vec{p}^{,\prime})$ the pure Coulomb amplitude, and $T_{SC}(\vec{p},\vec{p}^{,\prime})$ the strong amplitude in the presence of the Coulomb interaction—i.e., the strong amplitude calculated using Coulomb distorted wave functions instead of plane waves (see Refs.~\cite{Kong:1999sf,Albaladejo:2025kuv} for recent re-derivations). Neglecting the finite range of the strong interaction potential $V$, the amplitude $T_{SC}$ becomes purely $S$-wave. It can then be written in terms of the on-shell potential and the properties of the Coulomb $S$-wave function, as shown in Eqs.~(18) and  (20a, 20b) of Ref.~\cite{Albaladejo:2025kuv} for the  repulsive Coulomb case, and in Eqs.~(32) and (33) of the same reference for the  attractive one.

Here, we incorporate relativistic effects from the Klein–Gordon (KG) equation within a controlled set of approximations. In particular, consider a scalar particle of mass $m$ subject to an electrostatic Coulomb potential $V_C(r) \propto 1/r$, introduced through the time component of the covariant derivative. Neglecting the $V_C^2$ term, the resulting spatial KG wave function of the particle with momentum $p$ coincides with the solution $\psi(p,\vec r\,)$ of the non-relativistic Schr\"odinger equation~\cite{Hostler:1963zz}
\begin{equation}
    \left( -\vec\nabla^2+ 2\sqrt{m^2+p^2}\,V_C(r)-p^2\right)\psi(p,\vec r\,)=0
\end{equation}
The equation satisfied by two particles interacting through the Coulomb potential involves the reduced mass of the pair, which for $\Sigma_c\pi$ is well approximated by the pion mass. Hence, upon employing the relativistic resolvent implicit in the loop function introduced in the previous subsections, the results of Ref.~\cite{Albaladejo:2025kuv} yield
\begin{equation}
    T_{SC}(s)=\dfrac{C_\eta^2~e^{2i\sigma_0}}{V^{-1}(s)- G^{\rm Cb}_{\Sigma_c\pi}(s)},
\end{equation}
\begin{widetext}
where $C_\eta^2$ is the Sommerfeld factor and $\sigma_0$ is the $S$-wave Coulomb phase shift, given by
\begin{equation}
    C_\eta^2=\dfrac{2\pi \eta_p}{e^{2 \pi \eta_p}-1}, \quad \eta_p=\alpha e_{\rm eff}\,\dfrac{  \sqrt{\mu^2+p^2}}{p}, \quad
    e^{2i \sigma_0}=\dfrac{\Gamma(1+i\eta_p)}{\Gamma (1-i\eta_p)},\label{eq:coulrel}
\end{equation}
where $p=\lambda^{1/2}(s,M_{\Sigma_c}^2,m_\pi^2)/(2\sqrt{s})$ is the on-shell relativistic CM momentum introduced above, $\Gamma(z)$ denotes the Gamma function, $\mu$ is the reduced mass\footnote{Actually, when the relative motion is separated within the framework of the KG equation (see, for instance, Ref.~\cite{Koshelkin:2025xag}), an effective mass $m_1 m_2/\sqrt{s}$ appears instead of the usual reduced mass for the two-body system. For simplicity, and in the momentum region where Coulomb effects may be sizable for the $\Sigma_c\pi$ system, it is sufficiently accurate to approximate this effective mass by the reduced mass when accounting for such effects. } of the $\Sigma_c\pi$ system, $\alpha \simeq 1/137.036$ is the fine-structure constant, and $e_{\rm eff}$ is the product of the particle charges in units of the proton electric charge (for the $\Sigma_c^{++}\pi^+$ system, $e_{\rm eff}=+2$). For a repulsive Coulomb interaction, the relativistic Coulomb-dressed loop function $G^{\rm Cb}_{\Sigma_c\pi}(s)$ takes the form\footnote{In the case of an attractive Coulomb interaction, additional contributions from Coulomb bound states must be included, as discussed in \cite{Nieves:2024dcz} within the non-relativistic framework.}
\begin{equation}
     G^{\rm Cb}_{\Sigma_c\pi}(s)= \int \dfrac{d^3q}{(2\pi)^3} \dfrac{E(q)+\omega(q)}{2 E(q) \omega(q)} \dfrac{2M_{\Sigma_c}|\psi(\vec{q},\vec 0\,)|^2}{s-(E(q)+\omega(q))^2+i\varepsilon}, \qquad |\psi(\vec{q},\vec 0\,)|^2=\dfrac{2\pi \eta_q}{e^{2\pi \eta_q}-1} , \label{eq:CoulombDressedG} 
\end{equation}     
\end{widetext}
where $E$ and $\omega$ denote the relativistic energies of the $\Sigma_c$ and $\pi$, respectively, for an off-shell three-momentum $\vec q$. Within the zero-range approximation adopted for the strong interaction, the Coulomb dressing of the loop function introduces only the modulus squared of the Coulomb wave function evaluated at the origin, $|\psi(\vec q,\vec r=\vec 0)|^2$.

As previously mentioned, the $\Sigma_c^{++} \pi^+$ loop integral needs to be regularized. On the one hand, for the SU(4)-WT model, the $G^{\rm Cb}_{\Sigma_c\pi}$ loop integral is regularized numerically with an UV cutoff, as  in Refs.~\cite{Hofmann:2005sw,Mizutani:2006vq,Jimenez-Tejero:2009cyn}. On the other hand, for the $\Sigma_c\pi$[WT $\&$ CQM] model, we follow the one-subtraction scheme previously described. However, neither the divergent contribution of the loop function, $G^{\rm Cb}_{\Sigma_c\pi}(s_{\rm th})$, nor its finite part can be evaluated analytically with a cutoff regularization due to the presence of $|\psi(\vec{q},\vec 0\,)|^2$. Therefore, the subtraction procedure must be implemented numerically. The divergent contribution is now solved numerically using an auxiliary  sharp cutoff $\Lambda^{\rm aux}$. The finite part of the loop function, $\overline{G}^{\rm Cb}_{\Sigma_c\pi}(s)$, is defined through the difference
\begin{equation}
    \overline{G}^{\rm Cb}_{\Sigma_c\pi}(s)=\lim_{\Lambda^{\rm aux}\to \infty} \left[ G^{{\rm Cb},\Lambda^{\rm aux}}_{\Sigma_c\pi}(s)-G^{{\rm Cb},\Lambda^{\rm aux}}_{\Sigma_c\pi}(s_{\rm th}) \right],
\end{equation}
where both loop functions are evaluated with the same momentum auxiliary sharp cutoff $\Lambda^{\rm aux}$. Since the UV divergences of $G^{{\rm Cb},\Lambda}_{\Sigma_c\pi}(s)$ and $G^{{\rm Cb},\Lambda}_{\Sigma_c\pi}(s_{+})$ are identical, they cancel exactly in the subtraction, yielding a finite result in the limit $\Lambda^{\rm aux}\to\infty$. We check that when the Coulomb interaction is set to zero, i.e., in the limit $\alpha = 0$,  the loop function of Eq.~\eqref{eq:Gbar} is reproduced. In addition, the renormalization constant required by the $\Sigma_c\pi$ [WT $\&$ CQM] model is obtained from the Coulomb-dressed loop function at threshold, $G^{{\rm Cb},\Lambda}_{\Sigma_c\pi}(s_{\rm th})$, calculated using a sharp cutoff $\Lambda$ varied within the range $400$–$650$ MeV, and correlated with the parameter $d_c$, which determines the strength of the CQM vertex in Eq.~\eqref{eq:VCQM}.

The Coulomb-dressed $\Sigma_c^{++} \pi^+$ loop functions for the two models, computed with similar UV cutoffs ($\sim 650$ MeV), are shown as dashed lines in Fig.~\ref{fig:Gs}. As in the case without Coulomb interaction, the loop functions for the two schemes depart at higher CM total energies (or CM momenta) due to the different regularization prescriptions. Moreover, the repulsive Coulomb interaction between $\Sigma_c^{++}$ and $\pi^+$ reduces the attraction in the real part of the loop function and diminishes the magnitude of its imaginary part compared to the case without Coulomb effects. This latter behavior directly follows from Eq.~\eqref{eq:CoulombDressedG}
\begin{equation}
    {\rm Im}\left[\,\overline{G}_{\Sigma_c\pi}^{\rm Cb}(s)\right]= C^2_\eta\, {\rm Im}\left[\,\overline{G}_{\Sigma_c\pi}(s)\right]= -C^2_\eta\,\frac{M_{\Sigma_c}}{4\pi\sqrt{s}}\,p  
\end{equation}
as necessary to ensure the elastic unitarity of the full $S$-wave partial amplitude, which includes both $T_{SC}$ and the purely Coulomb contribution~\cite{Albaladejo:2024lam}.

\subsection{Strong isotensor phase shifts in presence of Coulomb  }

\begin{figure}[h!]
    \centering
    \includegraphics[width=\linewidth]{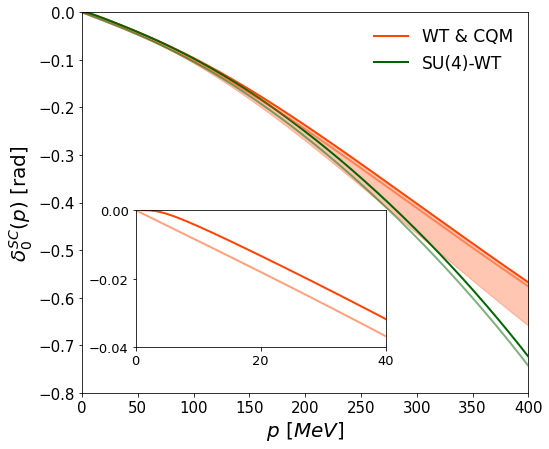}
    \caption{$\Sigma_c^{++} \pi^+$ $S$-wave strong phase shift in presence of the Coulomb interaction as a function of CM momentum. Darker orange (green) solid lines: $\Sigma_c\pi$ [WT $\&$ CQM] model with $\Lambda = 650~\mathrm{MeV}$ (SU(4)-WT model with $\Lambda = 653~\mathrm{MeV}$). Shaded band: variation of the $\Sigma_c\pi$ [WT $\&$ CQM] phase shift for UV cutoffs 400–650 MeV. Lighter orange and green lines: phase shifts without Coulomb effects.}
    \label{fig:PS_conC}
\end{figure}

We can now compute the $\Sigma_c^{++} \pi^+$ strong scattering amplitude in the presence of the Coulomb interaction. This is given by~\cite{Kong:1999sf,Albaladejo:2025kuv}
\begin{eqnarray}
    f^{-1}_{SC}(p)&=&C_\eta^2~p~(\cot{\delta_0^{SC}}-i)= -\dfrac{4 \pi \sqrt{s}}{M_{\Sigma_c}} \dfrac{C_\eta^2~e^{2i\sigma_0}}{T_{SC}\nonumber(s)} \nonumber\\ 
    &=& -\dfrac{4 \pi \sqrt{s}}{M_{\Sigma_c}}\left[V^{-1}(s)- G^{\rm Cb}_{\Sigma_c\pi}(s)\right] .
\end{eqnarray}
The $S$-wave strong phase shift $\delta_0^{SC}$ in the presence of the Coulomb interaction is shown as a function of the CM momentum in Fig.~\ref{fig:PS_conC}. 
As discussed above, the difference between the phase shifts predicted by the $\Sigma_c\pi$ [WT $\&$ CQM] and SU(4)-WT models for momenta above 200 MeV arises from their different regularization schemes. In addition, the repulsive Coulomb interaction slightly weakens the strong interaction at short distances, reducing the magnitude of the strong phase shift and bringing it closer to zero compared to the purely strong case, as illustrated in the inset for the $\Sigma_c\pi$ [WT $\&$ CQM] model.

From the scattering amplitude we can extract the $\Sigma_c^{++} \pi^+$ scattering length and effective range for the different models, using the effective range expansion as
\begin{equation}
    f^{-1}(p)=-\dfrac{1}{a_0}+\dfrac{1}{2}r_0p^2-ip+\cdots,
    \label{eq:ERE}
\end{equation}
for the purely strong case, and, when the Coulomb interaction is included~\cite{Kong:1999sf,Bethe:1949yr,Jackson:1950zz,vanHaeringen:1981pb,Konig:2012prq,Albaladejo:2025kuv}
\begin{align}
    f_{SC}^{-1}(p)&=\left( -\dfrac{1}{a_0^{SC}}+\dfrac{1}{2}r_0^{SC}p^2+...\right) \nonumber \\
    &-2\alpha e_{\rm eff} \sqrt{\mu^2+p^2} H(\eta_p), \label{eq:EREC}
\end{align}
where $ H(\eta_p)= \psi(i\eta_p)+1/(2i\eta_p)-\ln{i\eta_p}$ with $\psi(z)=\Gamma'(z)/\Gamma(z)$ the logarithmic derivative of the 
$\Gamma$-function, and  the principal
argument of the logarithm is taken between $]-\pi,\pi]$.

In Table~\ref{tab:ScattLen1}, we compile the $\Sigma_c^{++} \pi^+$ scattering lengths and effective ranges obtained within the SU(4)-WT and $\Sigma_c\pi$ [WT $\&$ CQM] models, using the corresponding UV cutoff values discussed above. We also provide the obtained low-energy parameters when Coulomb effects are included.
\begin{table}[h!]
\centering
\begin{tabular}{c|c|ccc}
\hline \hline Model & $\Lambda$[MeV] & $a_0^{(SC)}$ & $r_0^{(SC)}$  & $r_0^{(SC)}$  \\ 
                    &                &              &  (fit)           &  (der) \\\hline \hline
 WT & 400 & 0.194 & 7.77 & 7.86  \\
 WT+Coul. & 400 & 0.198 & 7.96 & 8.03 \\ \hline
 WT & 650 & 0.177 & 7.71 & 7.80 \\
 WT+Coul.& 650 & 0.181 & 7.91 & 7.97 \\ \hline
SU(4) & 653 & 0.177 & 8.26 & 8.34 \\ 
SU(4)+Coul. & 653 & 0.181 & 8.42 & 8.49 \\  \hline \hline
\end{tabular}
\caption{$\Sigma_c^{++} \pi^+$ scattering lengths and effective ranges in fermi units obtained within the $\Sigma_c\pi$ [WT $\&$ CQM] and SU(4)-WT models. For the $\Sigma_c\pi$ [WT $\&$ CQM] scheme, results are shown for the two limiting values of the UV cutoff. The effective ranges are extracted either from a quadratic fit to the scattering amplitude (fit) or from the derivative of the inverse scattering amplitude (der).}
\label{tab:ScattLen1}
\end{table}
In the absence of the Coulomb interaction, both models yield identical scattering lengths when a similar cutoff is employed. This is expected because, under these conditions, the loop function and the potential in this isospin channel coincide at threshold. Once the Coulomb interaction is included, the scattering length is only mildly modified. This limited effect is expected when the scattering length is small in magnitude, consistent with the absence of near-threshold bound or virtual states. In such cases, the scattering length is relatively insensitive to small variations in the interaction strength, unlike situations where a near-threshold state is present—such as in proton–proton scattering in the $S$-wave channel—where even tiny changes in the interaction can lead to sizable effects.

In fact, the strong scattering length, $a_0$, and the scattering length in the presence of the Coulomb interaction, $a_0^{\mathrm{SC}}$, are related through the difference between the standard and the Coulomb-dressed loop functions evaluated at threshold, 
\begin{eqnarray}
\frac{1}{a_0^{SC}}= \frac{1}{a_0}+\dfrac{4 \pi \sqrt{s_{\rm th}}}{M_{\Sigma_c}}\left[G_{\Sigma_c\pi}(s_{\rm th})- G^{\rm Cb}_{\Sigma_c\pi}(s_{\rm th})\right]
\end{eqnarray}
From Fig.~\ref{fig:Gs}, the difference between the loop functions at threshold is about $-1.9$  MeV, which translates into a change of approximately $\delta a_{\rm Cb}^{-1} = -0.1\ {\rm fm}^{-1}$ in the inverse scattering length when the Coulomb interaction is included. This corresponds to only a 2–3\% increase in the scattering length for the case under consideration. The effect of Coulomb repulsion on the strong scattering length would be much more significant if its inverse were comparable in magnitude to $\delta a_{\rm Cb}^{-1}$.

Regarding the effective range, two methods are employed for its extraction. The first consists of fitting the inverse scattering amplitude to a quadratic function of the momentum $p$ in the low-momentum region. The second method relies on computing the derivative of the inverse amplitude with respect to $p^{2}$ and evaluating it close to threshold. Both models yield similar values for the effective range. In each case, the inclusion of the Coulomb interaction leads to only a very small increase ($\lesssim 1\%$) in its value. This behavior can be understood as a consequence of the long-range nature of the Coulomb force, which modifies the wave function at large distances. As a result, the effective range—being sensitive to aspects of the interaction beyond the scattering length—receives additional contributions from the electromagnetic interaction that are suppressed by $\alpha$.

\section{$\Sigma_c^{++} \pi^+$ Correlation Function}
\label{sec:CorrelationFunction}

 In this section we proceed to calculate the $\Sigma_c^{++} \pi^+$ CF which can be calculated as (see, e.g.,~\cite{Koonin:1977fh,Pratt:1990zq,Fabbietti:2020bfg}):
\begin{align}
    &C(p)=\int_0^\infty d^3\vec{r} ~ S(r)
    \left| \phi(\vec{p},\vec{r})\right|^2 ,
    \label{eq:CF} 
\end{align}
where $\phi(\vec{p},\vec{r})$ denotes the scattering  full outgoing wave function of the hadron pair with CM momentum $\vec{p}$, and $S(r)$ is the source function describing the emission of the particle pair. In the following, we assume a Gaussian profile for the source.
\begin{equation}
    S(r)= (4\pi R^2)^{-3/2} \ \exp\left( -\dfrac{r^2}{4 R^2} \right),
\end{equation}
where $R$ defines the size of the source. With a partial wave decomposition of the wave function, and assuming a spherically symmetric source, Eq.~(\ref{eq:CF}) can be written as
\begin{align}
    &C(p)=\sum_{l=0}(2l+1)\int_0^\infty d^3\vec{r} ~ S(r)
    \left| \phi_l(\vec{p},\vec{r})\right|^2
    \label{eq:CF}.
\end{align}
Now, we can add and subtract the pure Coulomb wave function. Assuming that the strong interaction only affects the $S-$partial wave, we can decompose the CF
\begin{align}
    C(p)=&\int d^3\vec{r}~ {S}(r) ~ \left| \psi(\vec{r},\vec{p}) \right|^2+ \nonumber \\
    &\int d^3\vec{r}~  S(r) ~ \left( \left| \phi_0(r,p) \right|^2- \left| \psi_0(r,p) \right|^2 \right),
    \label{eq:CF2}
\end{align}
where $\psi(\vec{p},\vec{r})$ stands for the complete Coulomb wave function, $\psi_0(p,r)$ for its $S-$wave component, and $\phi_0(p,r)$ is the $S-$wave wave function including the strong and Coulomb interactions. 
The relativistic Coulomb wave functions are obtained from their standard non-relativistic counterparts by replacing the reduced mass $\mu$ in the definition of the parameter $\eta_p$ with $\sqrt{\mu^2+\vec{q}^{,2}}$, as discussed in Eq.~\eqref{eq:coulrel}. We follow the same normalization and conventions as in Ref.~\cite{Albaladejo:2025kuv}, where the explicit expressions can be found.

Asymptotically, for relative distances larger than the range of the strong interaction, the full radial $S$-wave wave function takes the form~\cite{Albaladejo:2025kuv}
\begin{align}
        &\phi_0^*\big|_{\rm asy}(r,\,p)=\frac{u^+_{\ell=0,p}(r)}{pr}+\dfrac{T_{SC}(s) ~\tilde{G}(s,r)}{e^{-\pi \eta_p/2} \Gamma(1+i \eta_p)}, 
\end{align}
where $u^+_{\ell=0,\,p}(r)$ represents the reduced Coulomb $S$-wave radial function, which in the conventions of Ref.~\cite{Albaladejo:2025kuv} reads $e^{\mathrm{i}\sigma_0}\,F_0(\eta_p, pr)$, and the Coulomb propagator $\tilde{G}(s,r)$ is defined as
\begin{widetext}
\begin{align}
    \tilde{G}(s,r)=& \int \dfrac{d^3q}{(2\pi)^3} \dfrac{2M_{\Sigma_c}(E(q)+\omega(q))}{2 E(q) \omega(q)}  \dfrac{e^{-\pi \eta_q/2}~\Gamma(1+i \eta_q)}{s-(E(q)+\omega(q))^2+i\varepsilon} \frac{\left[u^+_{\ell=0,\,q}(r)\right]^*}{qr} .
    \label{eq:Gtilde}
\end{align}
\end{widetext}

Let us focus on the asymptotic reduced full wave function,
\begin{equation}
u_0^*\big|_{\rm asy}(r,p)=r\,\phi_0^*\big|_{\rm asy}(r,p),
\end{equation}
which, unlike the exact solution, does not vanish at the origin but instead approaches the finite value $e^{i\sigma_0}C_\eta\,f_{SC}(p)$ at $r=0$. Substituting this expression into Eq.~\eqref{eq:CF2} leads to the Lednicky--Lyuboshitz approximation~\cite{Lednicky:1981su,Lednicky:1998} for the CF, including the relativistic corrections inherited from the BSE resolvent. Within the SU(4)-WT scheme, the exact reduced radial wave function is obtained by restricting the integration range of the off-shell momentum in Eq.~\eqref{eq:Gtilde}, implementing the same UV sharp cutoff $\Lambda$ used to determine the on-shell scattering amplitude, as first discussed in Ref.~\cite{Vidana:2023olz}. In fact, it is shown there that the half off-shell $T$-matrix, required to construct the wave function, can be obtained from the on-shell $T$-matrix by multiplying it by the same Heaviside function, $H(\Lambda - q)$, which is also introduced in the potential via the form factor $F_{\rm UV}$ to regularize the BSE in the UV regime.

Once the integration range in Eq.~\eqref{eq:Gtilde} is restricted, the resulting reduced radial wave function vanishes at $r=0$, and the corresponding CF becomes largely insensitive to moderate variations of this cutoff~\cite{Vidana:2023olz, Molina:2025lzw}. We have taken advantage of this feature and have similarly adopted this prescription to compute the $\Sigma_c\pi$ CF within the [WT $\&$ CQM] framework, which only determines the on-shell scattering amplitude.   The ambiguities in the [WT $\&$ CQM] CF arising when the momentum integration in Eq.~\eqref{eq:Gtilde} is cut between 400~MeV and 650~MeV are associated with the off-shell uncertainties discussed in \cite{Epelbaum:2025aan}. However, as noted above, these effects are very small in the present study and will not be addressed further; accordingly, we fix this cutoff at 650~MeV.

\begin{figure}[h!]
    \centering
    \includegraphics[width=\linewidth]{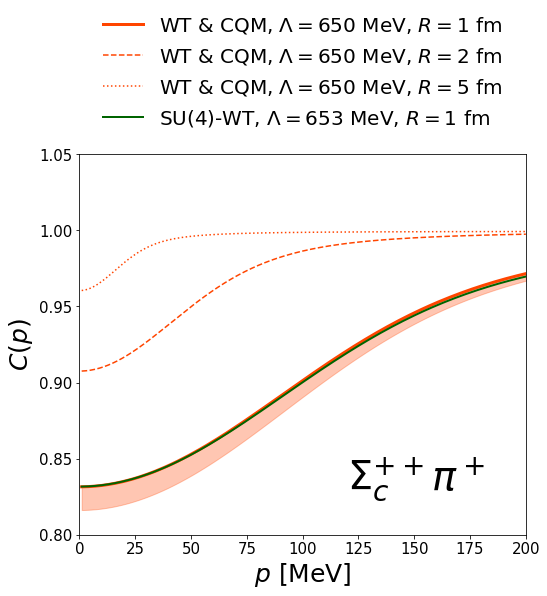}
    \caption{The $\Sigma_c^{++} \pi^+$ CF as a function of the CM momentum, calculated within the $\Sigma_c\pi$ [WT $\&$ CQM] (orange) and SU(4)-WT (green) models, including only the strong interaction, for a source radius of $R = 1~\mathrm{fm}$. For the $\Sigma_c\pi$ [WT $\&$ CQM] model with $\Lambda = 650~\mathrm{MeV}$ in the calculation of the on-shell amplitude, the CFs for source radii $R = 2~\mathrm{fm}$ (dashed) and $R = 5~\mathrm{fm}$ (dotted) are also shown. The band reflects the variations of the $\Sigma_c\pi$ [WT $\&$ CQM] CF, calculated for $R = 1~\mathrm{fm}$, arising from the ambiguities in the on-shell amplitude discussed in Figs.~\ref{fig:PS} and \ref{fig:PS_conC}.}
    \label{fig:CF_sinC}
\end{figure}

The $\Sigma_c^{++} \pi^+$ CF without the Coulomb interaction is shown in Fig.~\ref{fig:CF_sinC} as a function of the relative momentum, using both the $\Sigma_c\pi$ [WT $\&$ CQM] and the SU(4)-WT models. In this channel, the two schemes share the same irreducible potential, with their only difference arising from the regularization of the loop function that determines the on-shell scattering amplitude. For the $\Sigma_c\pi$ [WT $\&$ CQM] scheme, we employ two limiting cutoff values in the calculation of the on-shell amplitude, as discussed in Subsecs.~\ref{sec:i0} and \ref{sec:i2}. As no resonance or bound state is present in this channel, the resulting CF exhibits a smooth, featureless behavior. Nevertheless, a mild suppression below unity is observed at low relative momenta, reflecting the slightly repulsive nature of the strong $\Sigma_c\pi$ interaction, which can in turn be constrained through this effect. One should, however, expect Coulomb effects to significantly modify the CF in the low CM momentum region, as it must vanish at $p=0$ due to the electrostatic repulsion between the two positively charged particles.

\begin{figure}[h!]
    \centering
    \includegraphics[width=\linewidth]{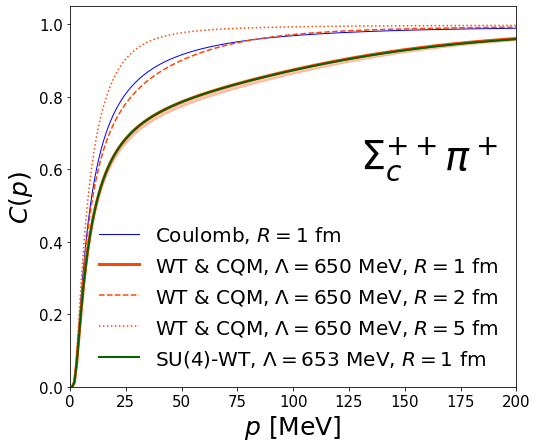}
    \caption{The same as in Fig.~\ref{fig:CF_sinC} but including the Coulomb interaction. In addition, we also display the CF for $R=1~\mathrm{fm}$ obtained when only the Coulomb interaction is included (blue curve).}
    \label{fig:CF_conC}
\end{figure}

\begin{figure}[h!]
    \centering
    \includegraphics[width=\linewidth]{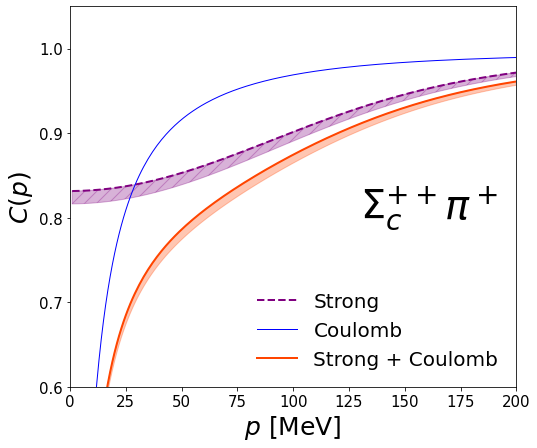}
    \caption{The same CFs for $R=1$ fm shown in Fig.~\ref{fig:CF_conC} are presented here, now also including the strong-interaction-only CF results (purple dashed curve and corresponding uncertainty band) previously displayed in Fig.~\ref{fig:CF_sinC}.}
    \label{fig:CF_conCzoom}
\end{figure}

In Figs.~\ref{fig:CF_conC} and \ref{fig:CF_conCzoom}, the $\Sigma_c^{++} \pi^+$ CFs corresponding to the scenarios discussed in Fig.~\ref{fig:CF_sinC}, now including Coulomb effects, are shown. For the $\Sigma_c\pi$ [WT $\&$ CQM] approach, the uncertainty band of the CF for $R=1~\mathrm{fm}$ becomes significantly narrower and is practically indistinguishable on the scale of Fig.~\ref{fig:CF_conC}, which is chosen to illustrate how rapidly the CF approaches zero as the momentum decreases. We observe that below 10 MeV the CF is overwhelmingly dominated by the repulsive Coulomb interaction between the  charged particles, which strongly suppresses any sensitivity to the details of the strong interaction. Consequently, once electromagnetic effects are included in this low-$p$ region, both the $\Sigma_c\pi$ [WT $\&$ CQM] and the SU(4)-WT strong-interaction models lead to very similar CFs. In Fig.~\ref{fig:CF_conCzoom} we focus on momenta above 25 MeV, where the effects of the strong interaction—potentially extractable from future measurements of the $\Sigma_c^{++}\pi^+$ CF—can be clearly appreciated. A certain sensitivity to the strong-interaction model is still observed, although distinguishing between the different scenarios may prove challenging.

In Figs.~\ref{fig:CF_sinC} and \ref{fig:CF_conC} we also display the $\Sigma_c^{++} \pi^+$ CF for different source radii. As the radius increases, the CF approaches unity, indicating that for sufficiently large source sizes the sensitivity to the details of the $\Sigma_c^{++} \pi^+$ interaction is progressively lost, as expected.

\section{$\Sigma_b^{+}\pi^{+}$ phase shift and Correlation Function}
\label{sec:CF_SbP}
 As anticipated in the Introduction, we also present here the results for the strong phase shift and the CF in the heavy-flavor counterpart $\Sigma_b \pi$ system in the bottom sector. In this case, we compare the predictions obtained in  Ref.~\cite{Lu:2014ina} to those derived from the [WT $\&$ CQM] scheme of Ref.~\cite{Nieves:2024dcz}. In Ref.~\cite{Lu:2014ina}, the lowest-lying isoscalar $\Lambda_b(5912)$ resonance in the $J^P = 1/2^-$ sector is dynamically generated by solving a coupled-channel BSE that includes the $\Sigma_b \pi$ and $\Xi_b' K$ channels. We employ the WT interaction given in Eq.~(\ref{eq:VWT}), with the corresponding coefficients listed in Table XXX of that work. The UV behavior of the BSE is regularized by means of a sharp cutoff, $\Lambda$, which is consistently applied to evaluate both the divergent and finite contributions to the $\Sigma_b \pi$ and $\Xi_b' K$ loop functions  using Eq.~\eqref{eq:gloop}.  We fix to $\Lambda = 1.62~\mathrm{GeV}$ in order to reproduce the $\Lambda_b(5912)$ bound state. In the original work of Ref.~\cite{Lu:2014ina}, the cutoff was $\Lambda = 2.17~\mathrm{GeV}$.  This difference is essentially due to the fact that Ref.~\cite{Lu:2014ina} uses an SU(3)-averaged pseudoscalar meson decay constant, $f_0 \sim 108~\text{MeV}$, about 15\% larger than $f_\pi$, in addition to approximations that assume the three-momentum of the baryon is small compared to its mass. This large UV scale, required to bring the molecular states down to the $\Lambda_b(5912)$ position, suggests—considering the arguments in Refs.~\cite{Guo:2016nhb, Albaladejo:2016eps}—the existence of additional relevant degrees of freedom (for instance, compact CQM states, as discussed in detail in Ref.~\cite{Nieves:2024dcz}) that are not included in Ref.~\cite{Lu:2014ina}.
 
In the isotensor $\Sigma_b^+\pi^+$ channel, whose dynamics we assume to be determined by the leading-order WT chiral interaction, we compare the predictions obtained using $\Lambda = 1.62~\mathrm{GeV}$ with those obtained using the more theoretically motivated range of $400$–$650~\mathrm{MeV}$, as done previously in the charm sector. This latter range of UV regulators provides a natural description of the heavy-quark-spin-symmetry doublet formed by the odd-parity $\Lambda_b(5912)$ and $\Lambda_b(5920)$ isoscalar states, with spins $1/2$ and $3/2$, respectively, within the [WT $\&$ CQM] approach introduced in Ref.~\cite{Nieves:2024dcz}.
\begin{figure}[h!]
    \centering
    \includegraphics[width=\linewidth]{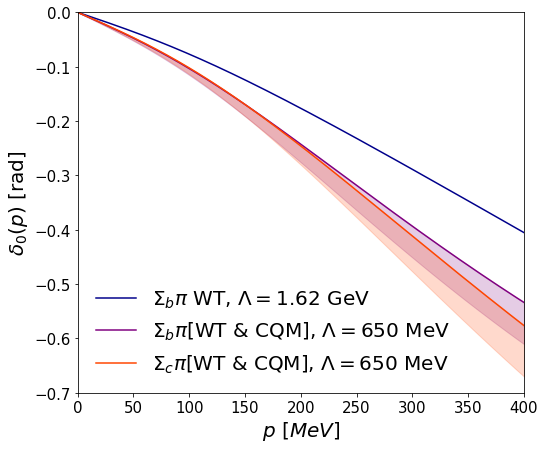}
    \caption{Strong isotensor $\Sigma_b\pi$ $S$-wave phase shift as a function of CM momentum. The purple and blue solid lines display the predictions of the $\Sigma_b\pi$ [WT $\&$ CQM] scheme with $\Lambda = 650~\mathrm{MeV}$ and of the WT model with $\Lambda = 1.62~\mathrm{GeV}$, respectively. The dark purple shaded band represents the range of phase shifts obtained within the $\Sigma_b\pi$ [WT $\&$ CQM] approach when the UV cutoff is varied between 400 and 650 MeV. The $\Sigma_c\pi$ [WT $\&$ CQM] strong phase shift in the charm sector, previously shown in Fig.~\ref{fig:PS}, is displayed in red.}
    \label{fig:PS_SbPi_sinC}
\end{figure}
In Fig.~\ref{fig:PS_SbPi_sinC}, we show the strong isotensor $\Sigma_b \pi$ $S$-wave phase shift as a function of the CM momentum. A significantly different behavior is observed depending on whether for the calculation of the on-shell amplitude a large cutoff of around $1.6~\mathrm{GeV}$ is used or the values adopted within the [WT $\&$ CQM] scheme are employed. Effects associated with the renormalization procedure also contribute to these differences. They arise from the use of a finite cutoff to evaluate ${\rm Re}\,\overline{G}_{\Sigma_b\pi}(s)$ in the calculation of Ref.~\cite{Lu:2014ina}. Such effects can remain non-negligible despite the relatively large value $\Lambda \sim 1.6~\mathrm{GeV}$, as a consequence of the sizable $\Sigma_b$ mass involved. They are absent in the [WT $\&$ CQM] scheme, where the cutoff is employed solely to determine $G_{\Sigma_c\pi}(s_{\rm th})$.

For reference, we also display in Fig.~\ref{fig:PS_SbPi_sinC} the [WT $\&$ CQM] strong phase shift previously obtained for the $\Sigma_c \pi$ system. At relative momenta below $200~\mathrm{MeV}$, the predictions for the charm and bottom sectors are nearly indistinguishable, as expected from HQFS, with only minor deviations appearing at higher momenta. In fact, both the potential in the BSE and the loop function are consistent with HQFS and are identical in the charm and bottom sectors, up to corrections proportional to $m_\pi$, the cutoff $\Lambda$, or the CM momentum $p$, all of which are suppressed by the heavy baryon mass $M_{\Sigma_{b,c}}$.  This approximate HQFS is also reflected in the [WT $\&$ CQM] $\Sigma_b\pi$ and $\Sigma_c\pi$ scattering lengths, for which those in the bottom sector differ by only about 2\% from the values compiled in Table~\ref{tab:ScattLen1} for the charm system. However, when the effective range expansion parameters are calculated for the large cutoff case, $\Lambda = 1.62~\mathrm{GeV}$, we find $a_0 = 0.14~\mathrm{fm}$, in reasonable agreement with the value listed in Table X of Ref.~\cite{Lu:2014ina}, and approximately 25\% smaller in magnitude than the results obtained using $\Lambda$ values in the range $400$–$650~\mathrm{MeV}$. The effective range is found to be $r_0 = 7.0~\mathrm{fm}$, which is also significantly different from the values previously obtained for the charm sector.  

We do not show the strong phase shift in the presence of the Coulomb interaction, $\delta_0^{SC}$, since the effective charge is $e_{\rm eff} = +1$, and the Coulomb effects are even smaller than in the $\Sigma_c\pi$ system, where the repulsion was twice as strong.
\begin{figure}[h!]
    \centering
    \includegraphics[width=\linewidth]{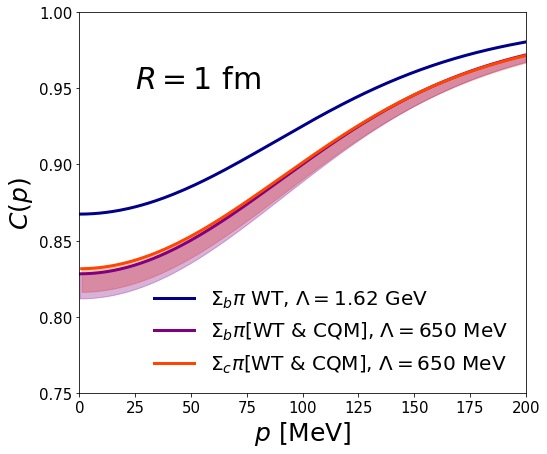}
    \caption{$\Sigma_b^+ \pi^+$ CF as a function of the relative momentum for a source radius $R = 1~\mathrm{fm}$. The results obtained with $\Lambda$ values in the range $400$–$650~\mathrm{MeV}$ for the calculation of the on-shell amplitude are shown by the purple curve and band, while those corresponding to $\Lambda = 1.62~\mathrm{GeV}$ are displayed in blue. In all cases, only the strong interaction is considered. The $\Sigma_c \pi$ CF in the charm sector, previously shown in Fig.~\ref{fig:CF_sinC} within the [WT $\&$ CQM] framework, is displayed in red.}
    \label{fig:CF_SbPi_sinC}
\end{figure}

The $\Sigma_b^+ \pi^+$ CF, including only the strong interaction and assuming a source radius $R = 1~\mathrm{fm}$, is shown in Fig.~\ref{fig:CF_SbPi_sinC} as a function of the relative momentum for the different UV regulators discussed above in the calculation of the phase-shifts. The differences observed in the phase shifts are clearly reflected in sizable variations of the predicted CF. Moreover, HQFS becomes manifest, as the charm and bottom sector [WT $\&$ CQM] CFs turn out to be practically identical. 
\begin{figure}[h!]
    \centering
    \includegraphics[width=\linewidth]{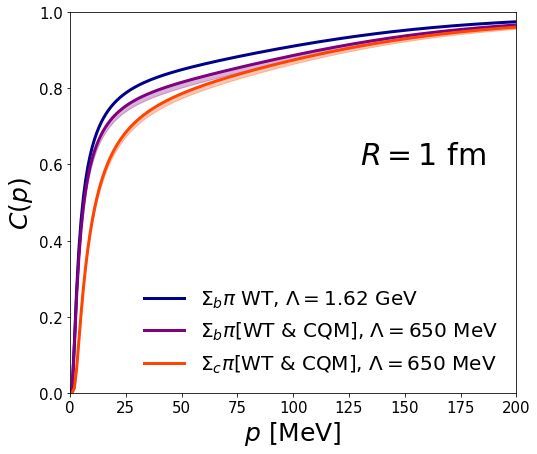}
    \caption{The same as is Fig.~\ref{fig:CF_SbPi_sinC}, but including the Coulomb interaction.}
    \label{fig:CF_SbPi_conC}
\end{figure}

As observed in the charm sector, Coulomb effects may obscure the sensitivity of the CF to the details of the strong interaction. Thus, in Fig.~\ref{fig:CF_SbPi_conC}, we display the $\Sigma_b^+ \pi^+$ CFs discussed above, now including Coulomb effects. While differences stemming from the strong interaction can still be distinguished in the intermediate-momentum region, the low-momentum behavior is dominated by Coulomb repulsion, even though the effective charge $e_{\rm eff}$ is half that in the charm sector.

\section{Summary}
\label{sec:Summary}

We first analyzed the $\Sigma_c^{++}\pi^+$ scattering amplitude and the corresponding strong phase shift within two different approaches, both adjusted to reproduce the properties of the lowest-lying odd-parity charm isoscalar spin-$1/2$ state, the $\Lambda_c(2595)$. The differences observed in this isotensor ($I=2)$ scattering amplitude—and consequently in the phase shift—at momenta above 200~MeV stem from the distinct UV regularization schemes adopted in the two frameworks.  Moreover, when the repulsive Coulomb interaction is included, the strong phase shift decreases compared to the scenario in which the electromagnetic interaction is switched off. Concerning the scattering length and effective range, we find a small (2\%) increase in both quantities once Coulomb effects are taken into account. 

We then computed the $\Sigma_c^{++}\pi^+$ CF without including the Coulomb interaction, analyzing the predictions obtained from the two models for different UV cutoff values. The CF remains below unity, reflecting the moderately repulsive character of the strong interaction, and displays a smooth behavior due to the absence of bound states or resonances in the $I=2$ sector. Subsequently, we incorporated Coulomb effects by employing relativistic Coulomb wave functions. Given the weak and non-resonant nature of the strong interaction, the CF is significantly influenced by the repulsive Coulomb interaction, which leads to nearly identical predictions from both realistic models for relative momenta $p \lesssim 50$~MeV. At higher momenta, however, the effects of the strong interaction become more pronounced and could, in principle, be extracted from future measurements of the $\Sigma_c^{++}\pi^+$ CF, as they are clearly visible in this region. In addition, a residual sensitivity to the underlying strong-interaction model persists, although experimentally distinguishing between the different scenarios may be challenging.

As pointed out at the end of the Introduction, predictions for the $\Sigma_c^{0}\pi^{-}$ channel correspond to those obtained for the $\Sigma_c^{++}\pi^{+}$ system when Coulomb effects are switched off. Therefore, the $\Sigma_c^{0}\pi^{-}$ CF—being free from Coulomb contributions—provides the most suitable observable to constrain the strong dynamics of the isotensor $\Sigma_c\pi$ system. This can be seen in Fig.~\ref{fig:CF_sinC}, where Coulomb effects are not included.
 
 We also calculated the phase shift for the heavy-flavor bottom partner channel $\Sigma_b^+\pi^+$ and confirmed the similarities  anticipated from HQFS. In this case, however, sizable differences arise in the $\Sigma_b^+\pi^+$ phase shift when the UV cutoff is taken to be very large—of the order of 1.6~GeV, as in Ref.~\cite{Lu:2014ina}—in order to reproduce the isoscalar resonance $\Lambda_b(5912)$ while including only chiral WT meson--baryon channels and neglecting CQM degrees of freedom. Regarding the $\Sigma_b^+\pi^+$ CF, the predictions obtained by including only the strong interaction exhibit a significant discriminating power between the different models. However, this sensitivity is partially washed out once the repulsive Coulomb interaction is taken into account.  In any case, regarding the CF presented in Fig.~\ref{fig:CF_SbPi_conC}, if it were to be measured, it could clearly favor/disfavor the results with an unrealistic UV cutoff of the order of 1.6 GeV. 

 The work of Ref.~\cite{Lu:2014ina} also explores the charm-quark sector and requires the introduction of a relatively large cutoff, above 1 GeV, in order to describe the isoscalar $\Lambda_c(2595)$ resonance. Consequently, if the $\Sigma_c^{++}\pi^+$ system were analyzed within that framework, the resulting phase shifts and CFs would differ appreciably from those presented in Secs.~\ref{sec:phase-length-range} and \ref{sec:CorrelationFunction}. For the sake of brevity, we have not included those results in the present study.

\begin{acknowledgments}
We  acknowledge support from the programs Unidades de Excelencia  Severo Ochoa CEX2023-001292-S and María de Maeztu CEX2020-001058-M, and from the projects PID2022-139427NB-I00 and PID2023-147458NB-C21 financed by the Spanish MCIN/AEI/10.13039/501100011033/FEDER, UE (FSE+), and by the Grant CIPROM 2023/59 of Generalitat Valenciana. 
\end{acknowledgments}

\newpage

\bibliography{refs}
\bibliographystyle{apsrev4-1}

\end{document}
%